# Resonant Interactions of Poincaré Waves in the Shallow Water Approximation

D. A. Klimachkov[a] and A. S. Petrosyan[a, b, *]

[a] *Space Research Institute, Russian Academy of Sciences, Moscow, Russia*
[b] *Moscow Institute of Physics and Technology (National Research University), Dolgoprudny, Moscow oblast, Russia*
*e-mail: apetrosy@cosmos.ru

**Abstract**—The paper develops a weakly nonlinear theory of Poincaré waves. The nondegeneracy of the Poincaré wave dispersion law leads to the presence of resonant interactions in perturbation theory. A study of the dispersion relation of Poincaré waves showed that three-wave interactions are absent in the quadratic nonlinear approximation. In this paper, a linear equation of the envelope is derived. A qualitative study of the dispersion law showed the existence of four-wave interactions of Poincaré waves. Equations of nonlinear interactions of four waves for the amplitudes of Poincaré waves are derived. The Manley–Rowe equations are obtained, which determine the distribution of energy and its transfer between interacting waves. The nonlinear dynamics of interacting waves is investigated. The saturation effect of Poincaré waves, which is important for geophysical hydrodynamics, has been predicted. An analytical solution is obtained that describes the saturation effect of Poincaré waves in time.

**Keywords:** nonlinear waves, resonant interactions, Poincaré waves, shallow water, hydrodynamics

## INTRODUCTION

The work is devoted to the weakly nonlinear theory of Poincaré waves. Poincaré waves have traditionally been studied in studies of large-scale rotating fluid flows on a flat boundary in the shallow water approximation (Pedlosky, 1987). Other types of waves in a rotating fluid, such as Kelvin waves and Rossby waves, are considered in the context of flow sphericity (Zeitlin, 2007). Among large-scale planetary waves, Kelvin waves and Poincaré waves are gravity waves modified by the presence of rotation (Majda et al., 1999; Bernard et al., 2008). Rossby waves are formed due to the latitude dependence of the Coriolis parameter in currents on a rotating sphere (Paldor et al., 2013; Raphaldini and Raupp, 2015).

Poincaré waves are slow, large-scale waves in a rotating heavy fluid. Since such waves belong to slow planetary hydrodynamic processes in which the condition of smallness of the Rossby number (the ratio of advective acceleration to the Coriolis acceleration) is fulfilled, therefore, in geophysical currents, hydrodynamics is determined by the Coriolis force, and the influence of the inertial force can be neglected (Dolzhansky, 2016). Poincaré waves shape the dynamics of the Earth's atmosphere (Williams et al., 2003) and planetary atmospheres (Peralta et al., 2014; Cho, 2008), also determine processes in the ocean (Nicolsky et al., 2018; Paldor et al., 2007).

In a wide range of works, large-scale flows of heavy fluid in geophysical hydrodynamics are described using the shallow water approximation (Vallis, 2006; Petviashvili and Pokhotelov, 1989). In this approximation, it is assumed that the characteristic lengths of the waves being studied are much greater than the height of the liquid layer. The classical shallow water equations are obtained from the complete three-dimensional system of hydrodynamic equations for a thin layer of incompressible fluid with a free boundary in a gravity field (Karelsky et al., 2012, 2013). The complete system is integrated along the vertical axis, taking into account the hydrostatic pressure distribution and the small layer height compared to the characteristic horizontal scale of the flow.

This paper is devoted to the development of the weakly nonlinear Poincaré wave theory (Didenkulova et al., 2011; Biancofiore et al., 2015). The dispersion law for Poincaré waves differs significantly from the dispersion law of gravitational waves in a nonrotating fluid. In the absence of rotation, the dispersion law of gravitational waves is degenerate, so the weak nonlinearity approximation does not allow resonant interaction of three waves in the first order of perturbation theory in small amplitude. However, even in this case, the effects of the next-order perturbation theory that determine the modulation of the beam of gravitational waves can be described in the weak nonlinearity approximation (Witham, 2010). In our case, the dispersion relation for Poincaré waves is nondegenerate due to rotation, which in the weak nonlinearity approximation can lead to resonant interactions. The work (Babin et al., 1997) shows the absence of resonant three-wave interactions for inertial-gravitational waves in the presence of rotation (Poincaré waves). The absence of resonant three-wave interactions for Poincaré waves is a consequence of the conservation of wave action (Falkovich, 1992; Falkovich and Medvedev, 1992; Glazman, 1996). A qualitative analysis of the shape of the dispersion surfaces of Poincaré waves shows the impossibility of fulfilling the synchronism conditions for three waves, and, as a consequence, the absence of three-wave resonances in the first order of perturbation theory for small amplitude. It should be noted that for magnetohydrodynamic flows in the shallow water approximation, the condition of three-wave resonances is satisfied, therefore, parametric instabilities of magneto-Poincaré waves arise due to the presence of three-wave interactions (Klimachkov and Petrosyan, 2016, 2017; Fedotova et al., 2023). In our work, we show that in this case resonant wave interactions exist in a higher order of perturbation theory, namely, the Poincaré wave dispersion law in the shallow water approximation allows four-wave interactions. The presence of rotation changes the nature of the dispersion relations of gravity waves, so that resonant interactions of Poincaré waves can become a determining factor in the nonlinear dynamics of waves in geophysical flows (Aubourg et al., 2017). Thus, the nonlinear dynamics of Poincaré waves is determined by four-wave interactions. We note an important paper (Onorato et al., 2009) showing that the nonlinear wave dynamics for classical quadratic equations in the Boussinesq approximation is also determined by resonant four-wave interactions, since in this case the dispersion law of gravitational waves is nondegenerate. In the section "Initial equations: Linear Waves" presents the initial shallow water equations for a rotating layer and writes out solutions in the form of linear Poincaré waves, which are necessary for further study of these equations. A qualitative study of the dispersion law is carried out, and it is shown that the resonant interactions of four Poincaré waves determine the nonlinear wave dynamics. In the section "Weakly nonlinear Poincaré wave theory: Envelope equation" the quadratic nonlinearity is investigated and a linear equation for dispersion-induced modulations is obtained. The amplitude equations of four-wave interactions of Poincaré waves are obtained. In the section "Four-wave interactions of Poincaré waves: Instabilities and saturation effect" instabilities in four-wave interactions of Poincaré waves are investigated and an analytical solution is found that describes the saturation effect of Poincaré waves. In some initial conditions, the nonlinear dynamics of interacting waves is described. Instability increments were found. Equations for energy transfer between interacting Poincaré waves are obtained.

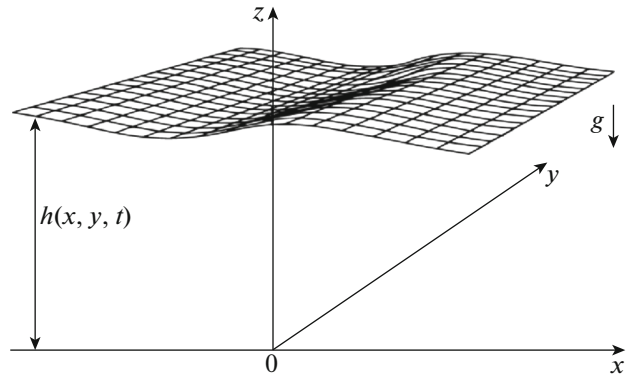

**Fig. 1.** Geometry of the problem.

## ORIGINAL EQUATIONS: LINEAR WAVES

To describe the flow of a thin rotating layer of liquid on a flat underlying surface with a free boundary in a vertical gravitational field, we will use hydrodynamic equations in the shallow water approximation (Fig. 1) (Karelsky et al., 2000a, 2000b):

$$\frac{\partial h}{\partial t} + \frac{\partial h v_x}{\partial x} + \frac{\partial h v_y}{\partial y} = 0, \tag{1}$$

$$\frac{\partial h v_x}{\partial t} + \frac{\partial \left(h v_x^2\right)}{\partial x} + \frac{\partial \left(h v_x v_y\right)}{\partial y} + gh\frac{\partial h}{\partial x} - fh v_y = 0, \tag{2}$$

$$\frac{\partial h v_y}{\partial t} + \frac{\partial \left(h v_x v_y\right)}{\partial x} + \frac{\partial \left(h v_y^2\right)}{\partial y} + gh\frac{\partial h}{\partial y} + fh v_x = 0. \tag{3}$$

In the system of Eqs. (1)–(3) $h$ is the layer height, $\mathbf{v}(v_x, v_y)$ is the horizontal velocity averaged over the layer height; $g$ is the acceleration of gravity, $f$ is the Coriolis parameter. The system of Eqs. (1)–(3) is the result of integrating three-dimensional hydrodynamic equations along the vertical axis $z$ from a flat underlying surface to a free surface $h$. The pressure distribution is assumed to be hydrostatic. The first equation of system (1) is a consequence of the continuity equation, Eqs. (2)–(3) are obtained as a result of averaging the equations for the horizontal impulse over the layer height. The system of Eqs. (1)–(3) is closed and is used to analyze linear waves and nonlinear interactions.

System (1)–(3) has a stationary solution $\mathbf{u}_0$, which is a rotating layer of liquid of height $h = h_0$ and with zero horizontal velocities $v_x = v_y = 0$. After linearization with respect to the stationary solution $\mathbf{u}_0 = (h_0, 0, 0)^T$ from the original system we get

$$\partial_t h + h_0 \partial_x v_x + h_0 \partial_y v_y = 0, \quad (4)$$

$$h_0 \partial_t v_x + g h_0 \partial_x h - f h_0 v_y = 0, \quad (5)$$

$$h_0 \partial_t v_y + g h_0 \partial_y h + f h_0 v_x = 0. \quad (6)$$

We are looking for a solution to system (4)–(6) in the form of a plane wave:

$$\begin{pmatrix} h \\ v_x \\ v_y \end{pmatrix} = \begin{pmatrix} h' \\ v'_x \\ v'_y \end{pmatrix} \exp(i(\omega t - (\mathbf{k}, \mathbf{r}))). \quad (7)$$

Substituting solution (7) into system (4)–(6), we obtain a system of linear equations:

$$\mathbf{H}_0 \begin{pmatrix} h' \\ v'_x \\ v'_y \end{pmatrix} = 0, \quad (8)$$

in which the linear operator $\mathbf{H}_0$ has the following form:

$$\mathbf{H}_0 = \begin{pmatrix} i\omega & -ik_x h_0 & -ik_y h_0 \\ -igk_x & i\omega & -f \\ -igk_y & f & i\omega \end{pmatrix}. \quad (9)$$

System (8) has nontrivial solutions when the condition is satisfied $\det \mathbf{H}_0 = 0$. We transform this condition to obtain the following dispersion relation for plane waves:

$$\omega^4 - \omega^2 (gh_0 k^2 + f^2) = 0. \quad (10)$$

The resulting dispersion equation has two solutions:

$$\omega_{1,2} = \pm \sqrt{C_0^2 k^2 + f^2} \quad (11)$$

is the dispersion relation for Poincaré waves, in which the expression with the sign "+" describes a wave propagating along a wave vector $\mathbf{k}$, and the expression with the sign "−" is a wave propagating in the opposite direction $\mathbf{k}$. In the dispersion law (11), the dynamics of plane waves in a rotating fluid, propagating in a layer of constant height, is determined by two restoring forces: gravity and the Coriolis force.

## WEAKLY NONLINEAR POINCARÉ WAVE THEORY: ENVELOPE EQUATION

To develop a weakly nonlinear theory from the general form of dispersion relations, we will determine the possibility of the existence of three-wave or four-wave resonances. For three interacting waves, the synchronism condition has the form:

$$\omega(\mathbf{k}_1) + \omega(\mathbf{k}_2) = \omega(\mathbf{k}_3), \quad \mathbf{k}_1 + \mathbf{k}_2 = \mathbf{k}_3. \quad (12)$$

In Fig. 2 sections of dispersion surfaces are shown by a plane $(\omega, k_x)$, corresponding to condition (12). The existence of intersection points in the obtained dispersion curves, and therefore in the dispersion surfaces, shows that there is a set of three waves that satisfy the synchronism condition. From Fig. 2 it is clear that dispersion curves cannot intersect, just like axisymmetric dispersion surfaces, since they are located in different regions of space, which means that dispersion relations for Poincaré waves exclude the presence of three waves satisfying condition (12). In the absence of three-wave synchronisms, the synchronism conditions for four interacting waves can be fulfilled:

$$\omega(\mathbf{k}_1) + \omega(\mathbf{k}_2) = \omega(\mathbf{k}_3) + \omega(\mathbf{k}_4), \\ \mathbf{k}_1 + \mathbf{k}_2 = \mathbf{k}_3 + \mathbf{k}_4. \quad (13)$$

In Fig. 3 sections of dispersion surfaces are shown by a plane $(\omega, k_x)$, corresponding to condition (13). The existence of intersection points of dispersion curves indicates the intersection of dispersion surfaces, which means that there are four Poincaré waves with dispersion law (10) that satisfy condition (13). Let us now write out the general solution of the linearized system, which is necessary for constructing a weakly nonlinear theory. In general, we write the solution $\mathbf{u}_1$ of the linear system of differential equations (4)–(6) as follows:

$$\mathbf{u}_1(T_0, X_0, Y_0) = \sum_{j=1}^{2} \int dk_x dk_y \mathbf{a}_j \alpha_j(\mathbf{k}) \\ \times \exp(i(\omega_j(\mathbf{k})T_0 - k_x X_0 - k_y Y_0)) + \text{c.c.}, \quad (14)$$

where $\omega_j(\mathbf{k})$ is determined by expressions (11), $\mathbf{a}_j$ are eigenvectors of a linear operator $\mathbf{H}_0$, and $\alpha_j(\mathbf{k})$ is the wave amplitude. From (8) we find the eigenvectors $\mathbf{a}_j$ of matrix $\mathbf{H}_0$. System (8) has nonzero solutions $\mathbf{a}_j \neq 0$ due to the fulfillment of condition (10).

$$\mathbf{a}_j = \begin{pmatrix} \omega_j(\mathbf{k}) g h_0 - i\omega_j(\mathbf{k}) k_x g h_0 f \\ i\omega_j(\mathbf{k}) f + k_x k_y g h_0 \\ -\omega_j^2(\mathbf{k}) + k_x^2 g h_0 \end{pmatrix} = \mathbf{a}_j(\mathbf{k}). \quad (15)$$

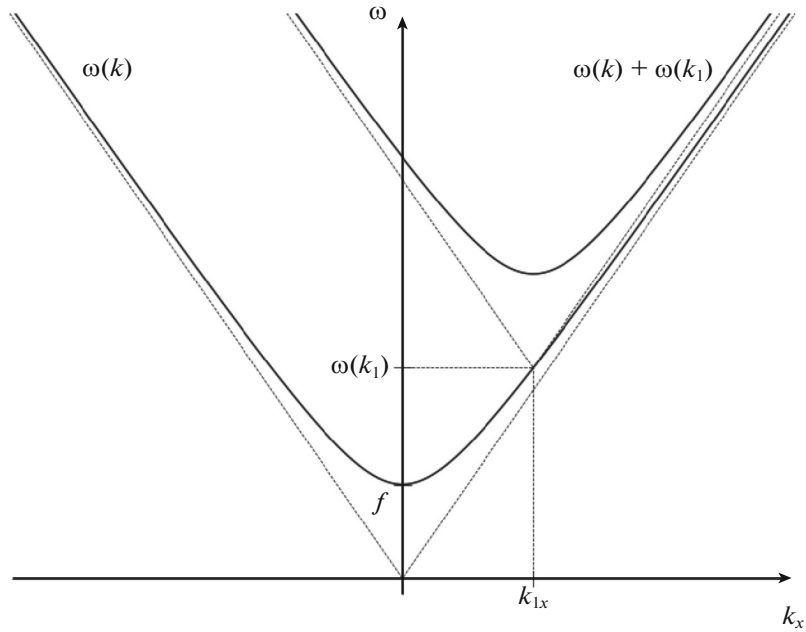

**Fig. 2.** Absence of three-wave interactions of Poincaré waves. The upper dispersion curve corresponds to the expression $\omega(k_1) + \omega(k)$, the lower dispersion curve corresponds to the expression $\omega(k)$. The dotted lines indicate the asymptotes of the dispersion curves.

In (15) $\omega_j = \omega_j(\mathbf{k})$, according to expressions (8).

For an analytical study of the nonlinear dynamics of Poincaré waves, we use the asymptotic method of multiscale expansions (Falkovich, 2011; Ostrovsky, 2014). The solution of system (4)–(6) is represented in the form of an asymptotic series in a small parameter $\varepsilon$:

$$\mathbf{u} = \mathbf{u}_0 + \varepsilon \mathbf{u}_1 + \varepsilon^2 \mathbf{u}_2 + \varepsilon^3 \mathbf{u}_3 + \ldots, \quad (16)$$

where $\mathbf{u}_0$ is a stationary solution against which small disturbances are considered, $\mathbf{u}_1$ is a solution of the linearized system (8), $\mathbf{u}_2$ is a correction describing the influence of quadratic nonlinearity, $\mathbf{u}_3$ is a correction describing the influence of cubic nonlinearity. Solving the equation for the following correction, we obtain a system of nonhomogeneous linear differential equations, on the left-hand side of which the linear operator $\mathbf{H}_0$ is valid for $\mathbf{u}_j$, and the right-hand side contains inhomogeneous terms, including previous corrections, which can cause resonance with the left-hand side of the system. Resonant terms lead to a linear growth of the solution, which on large scales violates the condition of asymptotic convergence of series (16) ($\varepsilon|\mathbf{u}_{j+1}| \ll |\mathbf{u}_j|$). Therefore, in order to eliminate the influence of resonant terms, we introduce the dependence of the wave amplitude on slow time and large spatial scales in the form $\mathbf{u}_j(T_{j+1}, X_{j+1}, Y_{j+1}) \exp(i(\omega T_j - k_x X_j - k_y Y_j))$. The condition that excludes the influence of resonant terms and ensures the convergence of the asymptotic series determines the evolution equation for a slowly changing amplitude. So, we move on from arguments $t, x, y$ to new arguments $(T_j, X_j, Y_j)$ so that with increasing index $j$ the transition to larger scales occurs, in accordance with the formulas

$$\begin{aligned}\frac{\partial}{\partial t} &= \frac{\partial}{\partial T_0} + \varepsilon \frac{\partial}{\partial T_1} + \varepsilon^2 \frac{\partial}{\partial T_2} + \ldots \\ \frac{\partial}{\partial x} &= \frac{\partial}{\partial X_0} + \varepsilon \frac{\partial}{\partial X_1} + \varepsilon^2 \frac{\partial}{\partial X_2} + \ldots \\ \frac{\partial}{\partial y} &= \frac{\partial}{\partial Y_0} + \varepsilon \frac{\partial}{\partial Y_1} + \varepsilon^2 \frac{\partial}{\partial Y_2} + \ldots\end{aligned} \quad (17)$$

We substitute into the original system (1)–(3) the solution in the form of a series (16) and expressions for partial derivatives (17) and write out the terms proportional to $\varepsilon^2$. As a result, we obtain the following equation:

$$\mathbf{H}_0(\mathbf{u}_2) = -\mathbf{P}_1(\mathbf{u}_1) - \mathbf{Q}_0(\mathbf{u}_1, \mathbf{u}_1), \quad (18)$$

where

$$\mathbf{P}_1(\mathbf{u}_1) = \begin{pmatrix} \frac{\partial h_1}{\partial T_1} + h_0 \frac{\partial v_{x1}}{\partial X_1} + h_0 \frac{\partial v_{y1}}{\partial Y_1} \\ h_0 \frac{\partial v_{x1}}{\partial T_1} + g h_0 \frac{\partial h_1}{\partial X_1} \\ h_0 \frac{\partial v_{y1}}{\partial T_1} + g h_0 \frac{\partial h_1}{\partial Y_1} \end{pmatrix}, \quad (19)$$

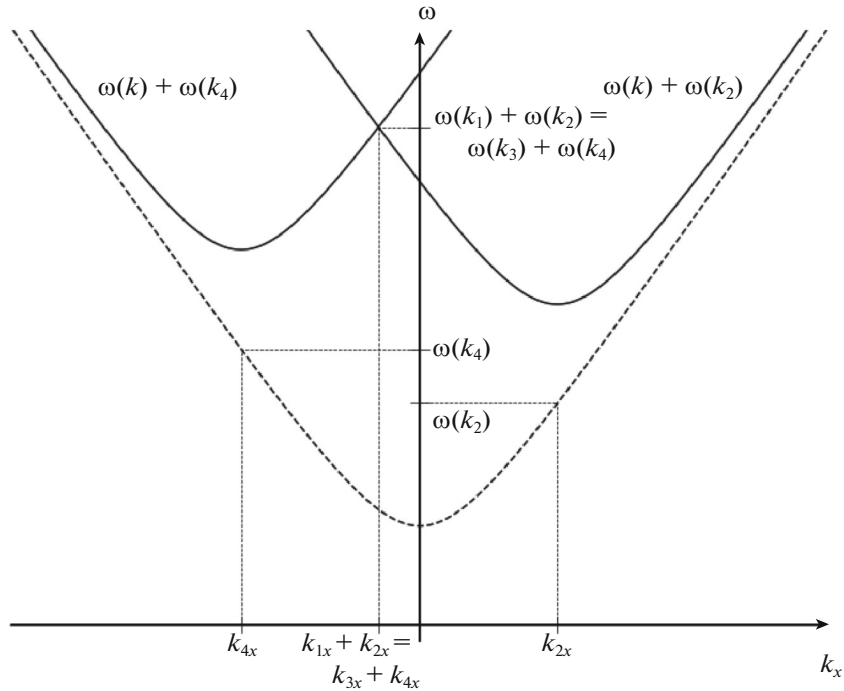

**Fig. 3.** The synchronism condition for four Poincaré waves is fulfilled. The left dispersion curve is defined by the expression $\omega(\mathbf{k}) + \omega(\mathbf{k}_4)$, the right one by expression $\omega(\mathbf{k}) + \omega(\mathbf{k}_2)$. The curve is indicated by a dotted line on which the points $(k_{4x}, \omega(\mathbf{k}_4))$ and $(k_{2x}, \omega(\mathbf{k}_2))$ lie.

$$\mathbf{Q}_0(\mathbf{u}_1, \mathbf{u}_1) = \begin{pmatrix} \dfrac{\partial h_1 v_{x1}}{\partial X_0} + \dfrac{\partial h_1 v_{y1}}{\partial Y_0} \\ \dfrac{\partial h_1 v_{x1}}{\partial T_0} + h_0 \dfrac{\partial (v_{x1}^2)}{\partial X_0} + h_0 \dfrac{\partial (v_{x1} v_{y1})}{\partial Y_0} + g h_1 \dfrac{\partial h_1}{\partial X_0} - f h_1 v_{y1} \\ \dfrac{\partial h_1 v_{y1}}{\partial T_0} + h_0 \dfrac{\partial (v_{x1} v_{y1})}{\partial X_0} + h_0 \dfrac{\partial (v_{y1}^2)}{\partial Y_0} + g h_1 \dfrac{\partial h_1}{\partial Y_0} + f h_1 v_{x1} \end{pmatrix}. \tag{20}$$

For the limited correction of the order $\varepsilon^2$ we will ensure that the compatibility (solvability) condition of system (18) is met. For the solvability of system (18), the orthogonality condition of the right-hand side of $(-\mathbf{P}_1(\mathbf{u}_1) - \mathbf{Q}_0(\mathbf{u}_1, \mathbf{u}_1))$ must be satisfied, the kernel of the operator $\mathbf{H}_0$.

In the previous section it was shown that the dispersion relation for Poincaré waves excludes three-wave resonant interactions, and hence the expression $\mathbf{Q}_0(\mathbf{u}_1, \mathbf{u}_1)$ does not contain secular terms.

Thus, in order to get rid of the secular terms on the right-hand side of system (18), we set $\mathbf{P}_1(\mathbf{u}_1) = 0$. The solution to this system of equations is a modulation of the solution to the linear problem in the form of a plane wave on scales $T_1, X_1, Y_1$ of many large periods and lengths of a plane wave:

$$\mathbf{u}_1(T_1, X_1, Y_1) = \sum_{j=1,2} \int dl_x dl_y \mathbf{b}_j \beta_j(\mathbf{l}) \\ \times \exp(i(\eta_j(\mathbf{l}) T_1 - l_x X_1 - l_y Y_1)) + \text{c.c.} \tag{21}$$

Here $\eta_{1,2} = \pm\sqrt{ghl^2}$, and the eigenvectors $\mathbf{b}_{1,2}$ of matrix $\mathbf{P}_1$, along which the oscillation is modulated, have the following form:

$$\mathbf{b}_{1,2} = \begin{pmatrix} \pm\sqrt{ghl^2} \\ gl_x \\ gl_y \end{pmatrix} = \mathbf{b}_{1,2}(\mathbf{l}). \tag{22}$$

The expressions (21) and (22) describe the dispersion spreading of the Poincaré wave packet on scales $T_1, X_1, Y_1$.

For further analysis of nonlinear effects of the order $\varepsilon^3$, we will find a complete solution for the quadratic correction $\mathbf{u}_2$. Thus, when conditions (21) and (22) are met, the term $\mathbf{P}_1(\mathbf{u}_1)$ is excluded from the system of equations. Then the system of Eqs. (18) for the correction $\mathbf{u}_2$ takes the form:

$$\mathbf{H}_0(\mathbf{u}_2) = -\mathbf{Q}_0(\mathbf{u}_1, \mathbf{u}_1). \tag{23}$$

The solution of this system will be the sum of the particular solution of the inhomogeneous system (23) and the general solution of the homogeneous system $\mathbf{H}_0(\mathbf{u}_2) = 0$. The quadratic correction $\mathbf{u}_2$ is satisfied with the condition $\varepsilon \mathbf{u}_2 \ll \mathbf{u}_1$. Thus, the solution of a homogeneous system $\mathbf{H}_0(\mathbf{u}_2) = 0$ contains plane waves that coincide with the solution of the linear problem (14).

Let us now find a particular solution for the inhomogeneous system (23). Note that the solution of the linear problem (14) in the form of a sum of several plane waves determines the type of nonlinear terms on the right-hand side of Eq. (23), namely, the nonlinear terms are proportional $\exp(i(\theta_j + \theta_k))$, where $\theta_j = \pm\omega_j T_0 \mp k_{xj}X_0 \mp k_{yj}Y_0$ are the phases of individual plane waves. Thus, we seek a particular solution of the inhomogeneous system (23) in the form:

$$\begin{pmatrix} h_2 \\ v_{x2} \\ v_{y2} \end{pmatrix} = \begin{pmatrix} h_2' \\ v_{x2}' \\ v_{y2}' \end{pmatrix} \exp(i(\theta_i + \theta_k)). \tag{24}$$

After substituting solution (24) into system (23), we divide both parts of the system by $\exp(i(\theta_j + \theta_k))$. Then we obtain the following system of linear equations:

$$\mathbf{L} u_2' = \mathbf{m}, \tag{25}$$

where

$$\mathbf{L} = \begin{pmatrix} \omega_j + \omega_k & -(k_{xj}+k_{xk})h_0 & -(k_{yj}+k_{yk})h_0 \\ -(k_{xj}+k_{xk})gh_0 & (\omega_j+\omega_k)h_0 & ifh_0 \\ -(k_{yj}+k_{yk})gh_0 & -ifh_0 & (\omega_j+\omega_k)h_0 \end{pmatrix}, \tag{26}$$

$$m = \begin{pmatrix} (k_{xj}+k_{xk})h_1 v_{x1} + (k_{yj}+k_{yk})h_1 v_{y1} \\ -(\omega_j+\omega_k)h_1 v_{x1} + (k_{xj}+k_{xk})\left(h_0 v_{x1}^2 + \frac{g}{2}h_1^2\right) + (k_{yj}+k_{yk})h_0 v_{x1} v_{y1} - ifh_1 v_{y1} \\ -(\omega_j+\omega_k)h_1 v_{y1} + (k_{xj}+k_{xk})h_0 v_{x1} v_{y1} + (k_{yj}+k_{yk})\left(h_0 v_{y1}^2 + \frac{g}{2}h_1^2\right) + ifh_1 v_{x1} \end{pmatrix}, \tag{27}$$

where $(h_1, v_{x1}, v_{y1})^T = \alpha \cdot \mathbf{a}$ are the components of the matrix eigenvector $\mathbf{H}_0$ (15), multiplied by the corresponding amplitudes of the linear solution $\alpha$. The solution to the resulting system (25) has the form $u_2' = \mathbf{L}^{-1} \mathbf{m}$. Thus, we write down the terms in the obtained for $u_2$, which can cause a resonant growth of the solution in the next order with respect to the small parameter $\varepsilon$:

$$\mathbf{u}_2(T_0, X_0, Y_0) = \sum_{j,k} (\mathbf{L}^{-1}\mathbf{m})_{j,k} \exp(i(\theta_j + \theta_k)). \tag{28}$$

## FOUR-WAVE INTERACTIONS OF POINCARÉ WAVES: INSTABILITIES AND SATURATION EFFECT

### Derivation of Equations of Four-Wave Interactions of Poincaré Waves

We investigate the effects of the next order in a small parameter $\varepsilon$. To do this, we substitute expressions (16) and (17) into the system (1)–(3) and write out the terms proportional to $\varepsilon^3$. Thus, we obtain a system in the third order in the small parameter $\varepsilon$ in the following form:

$$\begin{aligned}\mathbf{H}_0(\mathbf{u}_3) = &-\mathbf{P}_1(\mathbf{u}_2) - \mathbf{P}_2(\mathbf{u}_1) \\ &- \mathbf{Q}_0(\mathbf{u}_1, \mathbf{u}_2) - \mathbf{Q}_0(\mathbf{u}_2, \mathbf{u}_1) - \mathbf{R}_1(\mathbf{u}_1, \mathbf{u}_1) \\ &- \mathbf{S}_0(\mathbf{u}_1, \mathbf{u}_1, \mathbf{u}_1),\end{aligned} \tag{29}$$

where

$$\mathbf{P}_1(\mathbf{u}_2) = \begin{pmatrix} \dfrac{\partial h_2}{\partial T_1} + h_0 \dfrac{\partial v_{x2}}{\partial X_1} + h_0 \dfrac{\partial v_{y2}}{\partial Y_1} \\ h_0 \dfrac{\partial v_{x2}}{\partial T_1} + gh_0 \dfrac{\partial h_2}{\partial X_1} \\ h_0 \dfrac{\partial v_{y2}}{\partial T_1} + gh_0 \dfrac{\partial h_2}{\partial Y_1} \end{pmatrix}, \tag{30}$$

$$\mathbf{P}_2(\mathbf{u}_1) = \begin{pmatrix} \dfrac{\partial h_1}{\partial T_2} + h_0 \dfrac{\partial v_{x1}}{\partial X_2} + h_0 \dfrac{\partial v_{y1}}{\partial Y_2} \\ h_0 \dfrac{\partial v_{x1}}{\partial T_2} + gh_0 \dfrac{\partial h_1}{\partial X_2} \\ h_0 \dfrac{\partial v_{y1}}{\partial T_2} + gh_0 \dfrac{\partial h_1}{\partial Y_2} \end{pmatrix}, \quad (31)$$

$$\mathbf{Q}_0(\mathbf{u}_1, \mathbf{u}_2) = \begin{pmatrix} \dfrac{\partial h_1 v_{x2}}{\partial X_0} + \dfrac{\partial h_1 v_{y2}}{\partial Y_0} \\ h_0 \dfrac{\partial v_{x1} v_{x2}}{\partial X_0} + \dfrac{\partial h_1 v_{x2}}{\partial T_0} + h_0 \dfrac{\partial v_{x1} v_{y2}}{\partial Y_0} + gh_1 \dfrac{\partial h_2}{\partial X_0} - fh_1 v_{y2} \\ \dfrac{\partial h_1 v_{y2}}{\partial T_0} + h_0 \dfrac{\partial v_{x1} v_{y2}}{\partial X_0} + h_0 \dfrac{\partial v_{y1} v_{y2}}{\partial Y_0} + gh_1 \dfrac{\partial h_2}{\partial Y_0} + fh_1 v_{x2} \end{pmatrix}, \quad (32)$$

$$\mathbf{R}_1(\mathbf{u}_1, \mathbf{u}_1) = \begin{pmatrix} \dfrac{\partial h_1 v_{x1}}{\partial X_1} + \dfrac{\partial h_1 v_{y1}}{\partial Y_1} \\ \dfrac{\partial h_1 v_{x1}}{\partial T_1} + h_0 \dfrac{\partial v_{x1}^2}{\partial X_1} + h_0 \dfrac{\partial v_{x1} v_{y1}}{\partial Y_1} + gh_1 \dfrac{\partial h_1}{\partial X_1} \\ \dfrac{\partial h_1 v_{y1}}{\partial T_1} + h_0 \dfrac{\partial v_{x1} v_{y1}}{\partial X_1} + h_0 \dfrac{\partial v_{y1}^2}{\partial Y_1} + gh_1 \dfrac{\partial h_1}{\partial Y_1} \end{pmatrix}, \quad (33)$$

$$\mathbf{S}_0(\mathbf{u}_1, \mathbf{u}_1, \mathbf{u}_1) = \begin{pmatrix} 0 \\ \dfrac{\partial h_1 v_{x1}^2}{\partial X_0} + \dfrac{\partial h_1 v_{x1} v_{y1}}{\partial Y_0} \\ \dfrac{\partial h_1 v_{x1} v_{y1}}{\partial X_0} + \dfrac{\partial h_1 v_{y1}^2}{\partial Y_0} \end{pmatrix}. \quad (34)$$

In the resulting system (29) for the second correction we substitute the solution of the linearized system (4), (5), (6) in the form of a sum of four Poincaré waves satisfying the synchronism condition (13):

$$\mathbf{u}_1 = \varphi_1 \mathbf{a}(\mathbf{k}_1) \exp(i\theta_1) + \varphi_2 \mathbf{a}(\mathbf{k}_2) \exp(i\theta_2) \\ + \varphi_3 \mathbf{a}(\mathbf{k}_3) \exp(i\theta_3) + \varphi_4 \mathbf{a}(\mathbf{k}_4) \exp(i\theta_4) + \text{c.c.}, \quad (35)$$

where $\phi_j$ are the amplitudes of interacting Poincaré waves, $\theta_j = \omega(\mathbf{k}_j) T_0 - k_{xj} X_0 - k_{yj} Y_0$ are the phases of interacting waves ($j \in 1,2$), $\mathbf{a}$ is an eigenvector of the operator $\mathbf{H}_0$ (15), $\mathbf{a}^*$ is a complex conjugate vector. Now we define the terms on the right-hand side of the system of Eqs. (29) that interact with the operator on the left-hand side of $\mathbf{H}_0$, giving resonance. In the system of Eqs. (29), to eliminate the secular terms, we first assume $\mathbf{P}_1(\mathbf{u}_2) = 0$. This condition determines the dependence of the second quadratic correction on slow time and large spatial scales. Dependence of the linear solution on slow time and large spatial scales $\mathbf{u}_1(T_1, X_1, Y_1)$ (21) shows that the term $\mathbf{R}_1(\mathbf{u}_1, \mathbf{u}_1)$ does not provide resonant interaction with the operator on the left-hand side of $\mathbf{H}_0$, therefore, when considering resonant interactions in system (29), this term can be excluded. We find that system (29) is reduced to the following:

$$\mathbf{H}_0(\mathbf{u}_3) = -\mathbf{P}_2(\mathbf{u}_1) - \mathbf{Q}_0(\mathbf{u}_1, \mathbf{u}_2) \\ - \mathbf{Q}_0(\mathbf{u}_2, \mathbf{u}_1) - \mathbf{S}_0(\mathbf{u}_1, \mathbf{u}_1, \mathbf{u}_1). \quad (36)$$

The terms on the right-hand side of Eq. (36) include the found solutions of the problem of the first (14) and second (28) approximations with respect to the small parameter $\varepsilon$, resonant to the operator of the left-hand side of $\mathbf{H}_0$ in Eqs. (29). Thus, the compatibility condition (29) determines the dependence of the amplitudes of the linear Poincaré waves on the time scales and second-order coordinates $T_2, X_2, Y_2$ and the evolution of interwave interactions on these scales.

We will look for a solution to the system in the form of a sum of four interacting Poincaré waves (35), satisfying the synchronism condition (13). To satisfy the compatibility condition, as before, it is necessary to satisfy the condition of orthogonality of the right-hand side of system (36) to the kernel of the operator of the left-hand side. Let us write down the orthogonality condition for each of the eigenvectors of the operator $\mathbf{H}_0$ in the following form:

$$(\mathbf{z}, \mathbf{P}_2(\mathbf{u}_1) + \mathbf{Q}_0(\mathbf{u}_1, \mathbf{u}_2) \\ + \mathbf{Q}_0(\mathbf{u}_2, \mathbf{u}_1) + \mathbf{S}_0(\mathbf{u}_1, \mathbf{u}_1, \mathbf{u}_1)) = 0, \quad (37)$$

where $\mathbf{z}$ is the eigenvector of the operator $\mathbf{H}_0^*$. In expanded form, the compatibility condition has the following form:

$$\left( \begin{pmatrix} z_1 \\ z_2 \\ z_3 \end{pmatrix}, \begin{pmatrix} \dfrac{\partial}{\partial T_2}\sum_j a_{j1} + h_0 \dfrac{\partial}{\partial X_2}\sum_j a_{j2} + h_0 \dfrac{\partial}{\partial Y_2}\sum_j a_{j3} \\ h_0 \dfrac{\partial}{\partial T_2}\sum_j a_{j2} + gh_0 \dfrac{\partial}{\partial X_2}\sum_j a_{j1} \\ h_0 \dfrac{\partial}{\partial T_2}\sum_j a_{j3} + gh_0 \dfrac{\partial}{\partial Y_2}\sum_j a_{j1} \end{pmatrix} \right.$$

$$\left. + \begin{pmatrix} \dfrac{\partial}{\partial X_0}\xi_{12} + \dfrac{\partial}{\partial Y_0}\xi_{13} \\ \dfrac{\partial}{\partial T_0}\xi_{12} + \dfrac{\partial}{\partial X_0}(h_0\xi_{22} + \zeta_{122} + g\xi_{11}/2) + \dfrac{\partial}{\partial Y_0}(h_0\xi_{23} + \zeta_{123}) - f\xi_{13} \\ \dfrac{\partial}{\partial T_0}\xi_{13} + \dfrac{\partial}{\partial X_0}(h_0\xi_{23} + \zeta_{123}) + \dfrac{\partial}{\partial Y_0}(h_0\xi_{33} + \zeta_{133} + g\xi_{11}/2) + f\xi_{12} \end{pmatrix} \right) = 0, \quad (38)$$

where

$$\xi_{jk} = \sum_{l,m,n} \left\{ a_{lj}[(\mathbf{L}^{-1}\mathbf{m})_{m,n}]_k + a_{lk}[(\mathbf{L}^{-1}\mathbf{m})_{m,n}]_j \right\} + \text{c.c.}, \quad (39)$$

$$\zeta_{ijk} = \sum_{l,m,n} a_{li} a_{mj} a_{nk} + \text{c.c.} \quad (40)$$

In expressions (39) and (40), the summation is carried out over four interacting waves for each index $l, m, n$. For each component $r \in \{1, 2, 3, 4\}$ solutions (35), proportional to $\exp(i\theta_r)$, there is a set of works $a_{li}a_{mj}a_{nk}$ and $a_{lj}[(\mathbf{L}^{-1}\mathbf{m})_{m,n}]_k$, which will also give terms proportional to $\exp(i\theta_r)$. For each such set of resonant terms, condition (37) must be satisfied, in which $\mathbf{z} = \mathbf{z}_r$. Thus, from the compatibility condition (37) and the synchronism condition (13), after multiplying by the eigenvectors of the interacting waves, we obtain a system of equations describing the four-wave interactions of Poincaré waves in rotating hydrodynamic flows in the shallow water approximation in a gravity field with a free boundary, in the form of nonlinear equations with cubic nonlinearity:

$$\partial_{T_2}\varphi_1 + p_1\partial_{X_2}\varphi_1 + q_1\partial_{Y_2}\varphi_1 = \varphi_1 \sum_{j=1}^{4} s_{1j}|\varphi_j|^2 + f_1\varphi_2^*\varphi_3\varphi_4, \quad (41)$$

$$\partial_{T_2}\varphi_2 + p_2\partial_{X_2}\varphi_2 + q_2\partial_{Y_2}\varphi_2 = \varphi_2 \sum_{j=1}^{4} s_{2j}|\varphi_j|^2 + f_2\varphi_1^*\varphi_3\varphi_4, \quad (42)$$

$$\partial_{T_2}\varphi_3 + p_3\partial_{X_2}\varphi_3 + q_3\partial_{Y_2}\varphi_3 = \varphi_3 \sum_{j=1}^{4} s_{3j}|\varphi_j|^2 + f_3\varphi_1\varphi_2\varphi_4^*, \quad (43)$$

$$\partial_{T_2}\varphi_4 + p_4\partial_{X_2}\varphi_4 + q_4\partial_{Y_2}\varphi_4 = \varphi_4 \sum_{j=1}^{4} s_{4j}|\varphi_j|^2 + f_4\varphi_1\varphi_2\varphi_3^*. \quad (44)$$

Here $T_2, X_2, Y_2$ are slow time and coordinates, coefficients $a_i, b_i, s_i, f_i$, where $i \in \{1, 2, 3, 4\}$ are coefficients that depend only on the initial conditions and characteristics of the interacting waves:

$$p_r = \dfrac{h_0 z_{r1} a_{r2} + g z_{r2} a_{r1}}{(\mathbf{z}_r \cdot \mathbf{a}_r)}, \quad (45)$$

$$q_r = \dfrac{h_0 z_{r1} a_{r3} + g z_{r3} a_{r1}}{(\mathbf{z}_r \cdot \mathbf{a}_r)}, \quad (46)$$

$$s_{rt} = \dfrac{\mathbf{z}_r}{h_0(\mathbf{z}_r \cdot \mathbf{a}_r)}$$

$$\times \begin{pmatrix} ik_{rx}h_0\chi_{t12} + ik_{ry}h_0\chi_{t13} \\ -i\omega_r\chi_{t12} + ik_{rx}(h_0\chi_{t22} + \psi_{t122} + g\chi_{t11}/2) + ik_{ry}(h_0\chi_{t23} + \psi_{t123}) - f\chi_{t13} \\ -i\omega_r\chi_{t13} + ik_{rx}(h_0\chi_{t23} + \psi_{t123}) + ik_{ry}(h_0\chi_{t33} + \psi_{t133} + g\chi_{t11}/2) + f\chi_{t12} \end{pmatrix}, \quad (47)$$

$$f_r = \frac{\mathbf{z}_r}{h_0(\mathbf{z}_r \cdot \mathbf{a}_r)}$$
$$\times \begin{pmatrix} ik_{rx}h_0\nu_{12} + ik_{ry}h_0\nu_{13} \\ -i\omega_r\nu_{12} + ik_{rx}(h_0\nu_{22} + \mu_{122} + g\nu_{11}/2) + ik_{ry}(h_0\nu_{23} + \mu_{123}) - f\nu_{13} \\ -i\omega_r\nu_{13} + ik_{rx}(h_0\nu_{23} + \mu_{123}) + ik_{ry}(h_0\nu_{33} + \mu_{133} + g\nu_{11}/2) + f\nu_{12} \end{pmatrix}, \quad (48)$$

In expressions (47) and (48) the terms of the form $\chi_{tjk}, \psi_{tijk}$ and terms of the form $\nu_{tjk}, \mu_{tijk}$ obtained from $\xi_{jk}$ and $\zeta_{jk}$ and include only such works $a_{lj}[(\mathbf{L}^{-1}\mathbf{m})_{m,n}]_k$ and $a_{li}a_{mj}a_{nk}$, which are proportional to $\exp(i\theta_j)$ for each of the interacting waves.

### *Manley–Rowe Relation: Instabilities and Saturation Effect for Poincaré Waves*

In the resulting system (41)–(44), the following Manley–Rowe relation is satisfied for the energies of four interacting Poincaré waves (Craik, 1988):

$$\frac{\omega_m}{\theta_m}|\mathbf{a}_m|^2 - \frac{\omega_n}{\theta_n}|\mathbf{a}_n|^2 = \frac{\omega_m}{\theta_m}|\mathbf{a}_m(0)|^2 - \frac{\omega_n}{\theta_n}|\mathbf{a}_n(0)|^2, \quad (49)$$

where

$$m, n \in \{1, 2, 3, 4\}. \quad (50)$$

System (41)–(44) describes the interactions of four Poincaré waves satisfying the synchronism condition (13). Condition (49) describes the mechanism of energy transfer between interacting small-amplitude waves. In accordance with condition (49), an increase in the amplitude of the first of the interacting Poincaré waves leads to a decrease in the amplitude of the second interacting wave under conditions where the two remaining waves have a constant amplitude. The system of Eqs. (41)–(44) describes the dynamics of the change in the amplitude of one of the four interacting Poincaré waves in the case where the initial amplitude of the third and fourth waves is constant.

The rate of change of the amplitudes of interacting waves is obtained under the condition that the amplitudes of two of the four interacting waves are constant. To do this, we set the amplitudes in Eqs. (41)–(44) to be constants $\varphi_3 = \varphi_{30}$ and $\varphi_4 = \varphi_{40}$. Let also at the initial moment of time the amplitude of the second wave be much less than the amplitude of the first wave ($\varphi_2 \gg \varphi_1$). In this case, the system of equations of four-wave interactions (41)–(44) is linearized and takes the following form:

$$\frac{\partial \varphi_1}{\partial T_2} = i\varphi_1 \sum_{j=1}^{4} a_{1j}|\varphi_j|^2 + ih\omega_1\varphi_2^*\varphi_{30}^*\varphi_{40}^*\exp(-i\Delta\omega T_2), \quad (51)$$

$$\frac{\partial \varphi_2}{\partial T_2} = i\varphi_2 \sum_{j=1}^{4} a_{2j}|\varphi_j|^2 + ih\omega_2\varphi_1^*\varphi_{30}^*\varphi_{40}^*\exp(-i\Delta\omega T_2). \quad (52)$$

The resulting system (51), (52) has an exponentially growing solution:

$$\begin{pmatrix} \varphi_1 \\ \varphi_2 \end{pmatrix} = \begin{pmatrix} \varphi_{10} \\ \varphi_{20} \end{pmatrix} \exp(i\Gamma t) \quad (53)$$

with growth increment $\Gamma$:

$$\Gamma = \sqrt{|h^2\omega_1\omega_2|}\varphi_{30}\varphi_{40}. \quad (54)$$

Let us consider a special case of initial conditions in which one of the four interacting Poincaré waves is much smaller than the other three ($\varphi_{20}, \varphi_{30}, \varphi_{40} \gg \varphi_1$). Then at the initial moment the amplitude of each of the three waves $\varphi_{20}, \varphi_{30}, \varphi_{40}$ can be considered constant, while the reverse influence of the wave $\varphi_1$ on the pumping waves can be neglected. In this case, system (41)–(44) is linearized and takes the following form:

$$\frac{\partial \varphi_1}{\partial T_2} = i\varphi_1 \sum_{j=1}^{4} a_{1j}|\varphi_j|^2 + ih\omega_1\varphi_2^*\varphi_3^*\varphi_4^*\exp(-i\Delta\omega T_2). \quad (55)$$

The resulting Eq. (55) has an exponentially growing solution:

$$\varphi_1 = \varphi_{10}\exp(i\Gamma t) \quad (56)$$

with a growth increment $\Gamma$:

$$\Gamma = \sqrt{|h^2\omega_1^2|}\varphi_{10}\varphi_{20}\varphi_{30}. \quad (57)$$

Let us now consider the case when the energy of one of the interacting Poincaré waves is much greater than the energies of the other three waves. $\varphi_1 \gg \varphi_2, \varphi_3, \varphi_4$. Then, if we investigate the dependence of the wave amplitude only on time, Eq. (41) will take the form:

$$\partial_{T_2}\varphi_1 = \varphi_1 s_{11}|\varphi_1|^2. \quad (58)$$

The solution to this equation, in the case where the initial amplitude of the wave is $\varphi_1 = \varphi_{10}$ and the influence of the three remaining waves can be neglected, will be a decreasing function $\varphi_1(T_2)$ (Fig. 4):

$$\varphi_1(T_2) = \sqrt{\frac{\varphi_{10}^2}{1 - 2s_{11}\varphi_{10}^2 T_2}}. \quad (59)$$

Let us find the dependence of the amplitudes $\varphi_2, \varphi_3, \varphi_4$ on time, having considered the remaining equations of the system (42)–(44) and having accepted $\varphi_1 = \varphi_{10}$:

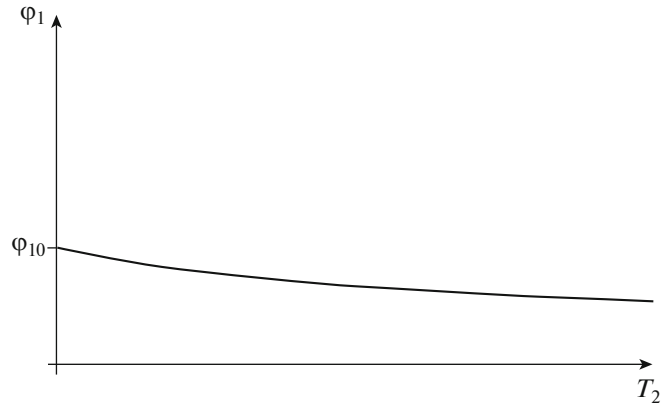

**Fig. 4.** Change in wave amplitude $\varphi_1$ depending on slow time $T_2$, defined in (17).

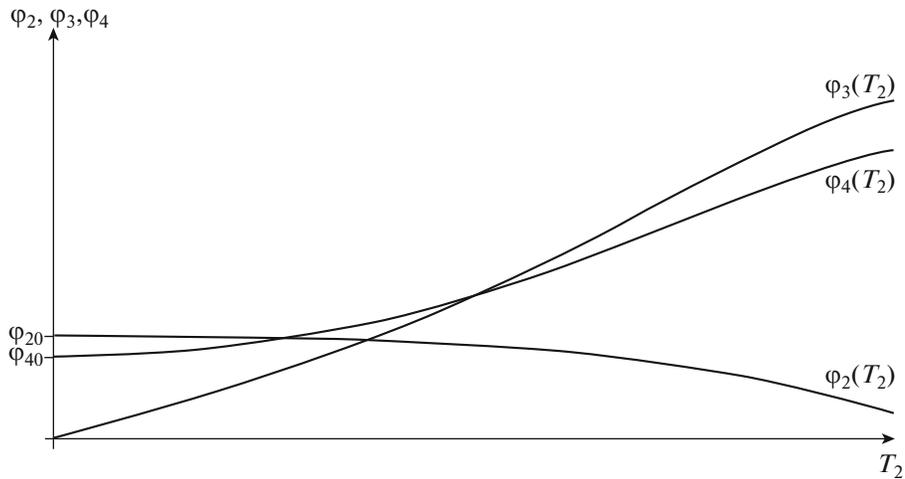

**Fig. 5.** Dynamics of dimensionless amplitudes of interacting Poincaré waves $\varphi_2, \varphi_3, \varphi_4$ depending on slow time $T_2$, defined in (17).

$$\partial_{T_2}\varphi_2 = \varphi_2 s_{21}|\varphi_{10}|^2 + f_2\varphi_{10}^*\varphi_3\varphi_4, \tag{60}$$

$$\partial_{T_2}\varphi_3 = \varphi_3 s_{31}|\varphi_{10}|^2 + f_3\varphi_{10}\varphi_2\varphi_4^*, \tag{61}$$

$$\partial_{T_2}\varphi_4 = \varphi_4 s_{41}|\varphi_{10}|^2 + f_4\varphi_{10}\varphi_2\varphi_3^*. \tag{62}$$

In the right-hand side of Eqs. (60)–(62), the first term determines the exponential growth of the amplitude with the increment:

$$\Gamma = s_{j1}|\varphi_{10}|^2, \tag{63}$$

and the second term describes the relationship between three growing interacting magnetogravitational waves. Let us find how the amplitudes of growing Poincaré waves change for some initial distribution of amplitudes $\varphi_{20}, 0, \varphi_{40}$ (Scott, 1977; Kadomtsev and Karpman, 1971). Dividing the first equation of the system (60) by the second (61), we obtain the relationship between the amplitudes $\varphi_2$ and $\varphi_3$:

$$|\varphi_2|^2 = \frac{f_2}{f_3}|\varphi_3|^2 + \varphi_{20}^2. \tag{64}$$

Similarly, we obtain the relationships for the remaining pairs of waves:

$$|\varphi_4|^2 = \frac{f_4}{f_3}|\varphi_3|^2 + \varphi_{40}^2, \tag{65}$$

$$|\varphi_4|^2 = \frac{f_4}{f_2}(|\varphi_2|^2 - |\varphi_{20}|^2) + \varphi_{40}^2, \tag{66}$$

The solution to the obtained system of Eqs. (64)–(66) are elliptic functions:

$$\varphi_2 = \varphi_{20} cd(\alpha T_2, w), \tag{67}$$

$$\varphi_3 = \sqrt{\frac{f_3}{f_4}} w \varphi_{40} sd(\alpha T_2, w), \qquad (68)$$

$$\varphi_4 = \varphi_{40} nd(\alpha T_2, w), \qquad (69)$$

where is the modulus of the elliptic function $w = 1 \Big/ \sqrt{1 + \frac{f_2}{f_3}\frac{\varphi_{30}^2}{\varphi_{20}^2}}$, and $\alpha = \varphi_{20}\sqrt{f_3 f_4}$. Thus, in the case under consideration, the three Poincaré waves with amplitudes much smaller than the amplitude of the fourth Poincaré wave $\varphi_2, \varphi_3, \varphi_4 \ll \varphi_1$, experience nonlinear interactions with saturation in accordance with solutions (67)–(69) (Fig. 5).

## CONCLUSIONS

The nonlinear wave dynamics of large-scale fluid flows in a gravity field in the presence of rotation is investigated. To study the linear and nonlinear dynamics of Poincaré waves in such flows, the shallow water approximation is used. It is shown that the dispersion relations for Poincaré waves exclude the possibility of three-wave interactions, and Poincaré waves in a rotating heavy fluid experience four-wave resonant interactions. In the first-order nonlinear correction in the wave amplitude, describing quadratic effects, a linear equation of envelope transfer is obtained, describing the modulation of propagating Poincaré waves on scales of the order of $\varepsilon k$, where $\varepsilon$ is a small parameter, and $k$ is the wave vector. In the next order of perturbation theory, describing cubic effects, a system of nonlinear equations is obtained for the amplitudes of four interacting Poincaré waves on scales of the order of $\varepsilon^2 k$, where $\varepsilon$ is a small parameter, and $k$ is the wave vector of the wave. Expressions for the interaction coefficients of Poincaré waves are found through the initial conditions of the problem and the characteristics of the interacting waves. The Manley–Rowe relation for the energies of four interacting Poincaré waves is found. It is shown that an increase in the amplitude of the first of the interacting waves leads to a decrease in the amplitude of the second interacting wave under conditions where the two remaining waves have a constant amplitude. The system of equations describes the growth of one of the four interacting waves in the case where the initial amplitude of the third and fourth waves is constant. The resulting system of nonlinear equations for the amplitudes of interacting waves is investigated in the approximation of a given pumping. It is shown that in the case where the amplitudes of three of the four interacting waves are much greater than the amplitude of the fourth wave, an exponential increase in the amplitude of the fourth wave occurs. Increments of such instabilities have been found. The saturation effect of Poincaré waves as a result of four-wave interactions has been discovered.

## FUNDING

The work was financed by funds from the Planet theme of the Space Research Institute of the Russian Academy of Sciences. No additional funding was received to conduct or supervise this study.

## CONFLICT OF INTEREST

The authors of this work declare that they have no conflicts of interest.